\documentclass{article}
\usepackage{arxiv}
\usepackage{cite}
\usepackage[utf8]{inputenc} 
\usepackage[T1]{fontenc}    
\usepackage{hyperref}       
\usepackage{url}  
\usepackage{graphicx}
\usepackage{booktabs}       
\usepackage{amsfonts}       
\usepackage{nicefrac}       
\usepackage{microtype}      
\usepackage{lipsum}
\usepackage{caption}
\usepackage[bottom]{footmisc}
\usepackage{authblk}
\usepackage{amsmath,amssymb,amsfonts}
\usepackage{float}
\usepackage{subfig}
\title{Transfer Learning for Automated Segmentation of Prostate Whole Gland and Transition Zone in Diffusion Weighted MRI}

\author[a, b]{Saman Motamed}
\author[c]{Isha Gujrathi}
\author[b]{Dominik Deniffel}
\author[d]{Anton Oentoro}
\author[a,b,d]{Masoom A. Haider}
\author[a,d,e,f *]{Farzad Khalvati}
\affil[a]{Institute of Medical Science, University of Toronto, Toronto, Canada}
\affil[b]{Lunenfeld-Tanenbaum Research Institute, Sinai Health System, Toronto, Canada}
\affil[c]{Brigham and Women’s Hospital, Boston, United States}
\affil[d]{Department of Medical Imaging, University of Toronto, Toronto, Canada}
\affil[e]{Department of Mechanical and Industrial Engineering, University of Toronto, Toronto, Canada}
\affil[f]{The Hospital for Sick Children, Toronto, Canada}
\begin{document} 
\maketitle
\begin{abstract} 
\textbf{Purpose:} Convolutional neural networks which have shown to be successful in various medical image analysis tasks including segmentation of prostate whole gland (WG) and transition zone (TZ) are typically sensitive to the variations in imaging parameters. These variations render performance of a trained model on new images from a different cohort with different acquisition parameters. In this work, we propose a transfer learning method based on a modified U-net architecture to improve segmentation results of images of a new cohort (Target) by using learnt features from a trained model on a bigger cohort (Source).
\textbf{Approach:} We explore the effect of the size of subset of target dataset used for fine-tuning the pre-trained CNN on the overall segmentation accuracy. We also propose a modification to the widely used Dice score coefficient (DSC) to improve performance in cohorts with small size. 
\textbf{Results:} Our results show that with a fine-tuning data as few as 30 patients from the target domain, the proposed transfer learning-based algorithm can reach mean DSC of 0.80 for both prostate WG and TZ segmentation where no transfer learning results in DSC of 0.65 for prostate WG and 0.51 for TZ. Using a fine-tuning data of 115 patients from the target domain, mean DSC of 0.85 and 0.84 are achieved for segmentation of WG and TZ, respectively, in the target domain compared to DSC of 0.71 without transfer learning.
\textbf{Conclusion:} Transfer learning improves segmentation results in cohorts with smaller size by taking advantage of learnt features from bigger cohorts. Training size and the corresponding performance can guide health care practitioners in labeling the right amount of data for the task of transfer learning. 
\end{abstract}
\section{Introduction}
Prostate cancer remains the second most commonly diagnosed cancer and one of the leading causes of cancer death for men in developed countries~\cite{39}. Magnetic resonance imaging (MRI) has become a successful alternative to the clinical standard transrectal ultrasound (TRUS) for cancer detection, localization, staging, biopsy guidance, and focal therapy \cite{41}. More recently, multi-parametric (mp)-MRI has played an increasingly important role in prostate cancer assessment. In addition to the excellent soft tissue contrast, mp-MRI can provide metabolic, diffusion, and perfusion information of prostatic tissue that improves the identification of possible cancerous regions. As the use of mp-MRI increases in clinical practice, as a part of the clinical decision support system, automated prostate cancer detection and segmentation can help radiologists interpret images faster and more accurately. An important step in this process is automated segmentation of prostate whole-gland (WG) and transition zone (TZ) given the tedious nature of manual contouring. Accurate segmentation of prostate and related anatomic structures is an essential task for a number of clinical workflows including radiation treatment planning. In addition, prostate volume has been shown to be a clinical factor in prostate cancer diagnosis~\cite{Roobol2012}.

\par Computer-aided detection (CAD) algorithms proposed for automated detection of prostate cancer rely on the segmentation of the prostate gland statistical processing step~\cite{Cameron2016, Chung2015, 30, Khalvati2018, 29, 28,DBLP:journals/corr/abs-1905-13145, meyer2019towards, jensen2019prostate}. Recently, deep convolutional networks (CNNs) have led to a series of breakthroughs in the field of computer vision. CNN models such as U-net~\cite{3} based architectures have been explored and shown prominent results in medical image segmentation. It is of high relevance to perform prostate WG and TZ segmentation on MR modalities such as T2-Weighted (T2w) and Diffusion Weighted Imaging (DWI). T2w MR images have higher spatial resolution compared to DWIs and hence easier for CNNs to achieve high segmentation results. DWI however can reveal the internal prostatic anatomy, prostatic margins, and the extent of prostatic tumors and are used in functional assessment of the prostate tissue being closely related to cancer biology~\cite{2, 3, 102}. In a study by Haider et al \cite{100}, combining T2w and diffusion-weighted MRI has shown improved results in localization of prostate cancer compared to analysis on T2w and DWI separately. Another study \cite{101} found that DWI is the primary determining sequence (dominant technique) for clinical cancer diagnosis assessment in prostate peripheral zone. Due to the duration of multiparametric prostate MRI protocols, minor movements by the patients are very common, leading to inaccurate co-registration between T2w and DWI. Any co-registration mismatch between the sequences can impair the performance of segmentation models that try to segment T2w images first and do registration between T2w and DWI to segment prostate in DWI. For these reasons, our focus on this work will be on segmentation on prostate in DW-MRIs without using T2w images. DW images, however, are subject to variation between different cohorts due to variance in acquisition parameters and scanner manufacturing differences.
\par
\begin{figure}
    \centering
    \subfloat[b-100 DWI sample from Source Domain]{{\includegraphics[width=4cm, height = 4cm]{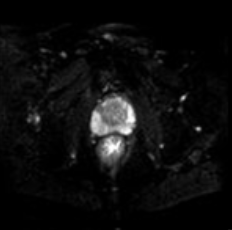} }}%
    \qquad
    \subfloat[b-100 DWI sample from Target Domain]{{\includegraphics[width = 4 cm, height = 4 cm]{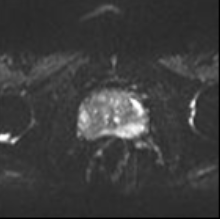} }}%
    \caption{b-100 images of prostate in two cohorts}%
    \label{fig:example1}%
\medskip
\small
\end{figure}

 Image variability limits the segmentation performance of CNNs trained on DWI of a given scanner (source domain) when applied to DWI acquired by a different scanner (target domain). DWI are acquired with different b-values, which reflect the strength and timing of the gradients used to generate the images~\cite{38}. Fig ~\ref{fig:example1} shows DWI of prostate from two different cohorts with b-value of $100s/mm^2$. 
 

As it can be seen from Fig~\ref{fig:example1}, there is distinguishable difference in quality and resolution of b-100 images from these two cohorts.

\par The task of acquiring medical imaging data and the process of manual contouring of images by radiologists can be cumbersome and expensive. Thus, it is imperative to be able to use all available labeled data from different cohorts and scanners, and transfer images and labels from one (source) cohort to the target cohort so that training a segmentation algorithm for the target cohort becomes possible with a minimal amount of labeled images. Data variability between cohorts and scanners remains a challenge for transferring learned knowledge from one cohort to another. Transfer learning is a promising approach to close the gap caused by data variability in image acquisition~\cite{42}.

\par Transfer learning is a deep learning technique that enables harnessing a neural network that has been trained on one domain to be applied to another domain. Transfer learning is useful when there is insufficient data for training a neural network on a new (target) domain while there is sufficient training data in another domain (source domain) that can be transferred to the task at hand. The lack of enough data is a major challenge in medical imaging and DWI variability  can benefit from a transfer learning approach. 

In this work, we propose a deep transfer learning approach for segmentation of prostate WG and TZ based on a modified U-net architecture. A deep learning architecture pretrained on a relatively large cohort (n=533) was fine-tuned using a small dataset from the target cohort and tested on the remaining data of the target cohort (n=33). The source and target cohorts were acquired from two different hospitals using different MRI scanners (Philips scanner for source cohort and Siemens scanner for target cohort). We explored the increase in accuracy of prostate WG and TZ segmentation, using different target domain dataset sizes for fine-tuning (n=8-115). Although transfer learning \cite{4, 20, 21, 42} has been explored extensively in deep learning for different CNN architectures for detection tasks, it has not been thoroughly investigated for segmentation tasks using U-net architecture. A recent work ~\cite{105} reported limited work on transfer learning using U-net for segmentation of breast in DWI but the detail of the transfer learning architecture and data split was not provided. Another work proposed an unsupervised domain adaptation ~\cite{104} for segmentation using an architecture similar to U-net. This is different to our transfer learning method where a minimum number of labeled cases from target domain is required. Although applied to different organs, the segmentation accuracy of our transfer learning algorithms is significantly higher than that reported by ~\cite{104}.

\par Additionally, we propose a modification to a widely used loss function in order to improve the performance of our model. Mean Dice score coefficient (DSC)~\cite{8} with values between 0 (no overlap of prediction and ground truth) and 1 (perfect segmentation prediction) has been used to evaluate the performance of medical image segmentation algorithms. In order to achieve better performance in segmentation tasks, CNN architectures such as VGG-16 have been used~\cite{1, vgg} as a preprocessing step to first detect prostate and only then train the segmentation architecture on images containing prostate~\cite{1}. With limited amount of data, however, training more architectures leads to loss of information over each model. To explore the effect of different rewards for prostate and non-prostate images and their corresponding segmentation predictions, we made a modification to the conventional DSC loss function and studied the effect of the modification on overall DSC, sensitivity, specificity and precision of each model's performance.

\section{Related Work}

Over the past few years, as the importance of prostate MRI segmentation grows, several segmentation methods have been proposed for the task of prostate zonal segmentation, including registration-based methods~\cite{Khalvati2015c, Khalvati2013,10, 11, 12}. Recently, deep learning methods have made advances in semi- and fully automatic segmentation of medical images including prostate~\cite{ 3, 1, 18, 19, jensen2019prostate, meyer2019towards}.
\par Recently, CNNs~\cite{31} have been a ground-breaking addition to existing segmentation algorithms, which are dominating the field of Computer Vision. CNNs have been responsible for significant advancements in tasks such as object classification~\cite{32} and object localization~\cite{33}, and the continuous improvements to CNN architectures are bringing further radical progresses~\cite{34,35, 36}. Semantic Segmentation tasks have also been revolutionized by CNNs. The 'U-net' architecture, proposed by Ronnenberger~\cite{3} was a great success for the task of medical image segmentation and has been used as a skeleton architecture in many studies~\cite{2, 1, 22}. The structure of U-net comprises of an encoder and a decoder network. Furthermore, the corresponding layers of the encoder and decoder networks are connected by skip connections prior to a pooling and subsequent to a deconvolution operation, respectively. U-net has been showing promising potential in segmenting medical images, even with a scarce amount of labeled training data. We have used this architecture as the base model and have proposed a transfer learning based segmentation algorithm for prostate WG and TZ.

\par With the rise of deep learning models which heavily depend on data size in order to be trained sufficiently and perform tasks with high accuracy, transfer learning methods are specially of interest in the field of medical imaging where data scarcity and MR variation between different institutes and image acquisition protocols significantly affect the performance of deep learning approaches. There is however, a lack of studies in exploring the amount of data needed in order to perform transfer learning from a source domain to a target domain. This is especially of high importance since the task of contouring medical images by radiologists is expensive, time consuming, and suffers from human-error and inter- and intra-user variability.

\par In a study by Ghafoorian et. al~\cite{4}, the effect of transfer learning on brain lesion segmentation of MR images were explored. They trained a CNN on legacy MR images of brain and evaluated the performance of the domain-adapted network on the same task with images from a different domain, reporting the DSC increase based on different dataset sizes used as the target domain. As of now, studies that explore the data size effects on different tasks such as prostate segmentation are very limited in the literature, yet it is of high importance to know how many labeled images are needed in order to efficiently extend a study from one cohort to another.

In this work, we propose an architecture with a transfer learning based approach for the task of prostate WG and TZ segmentation. The study can also act as a guide for any future work by measuring the amount of data needed for the task of prostate segmentation via transfer learning, by comparing final accuracy results based on multiple training size instances.

\section{Data and Methods}

\subsection{Datasets}
To perform the task of Prostate WG and TZ segmentation, and extending our trained model to different DWI domains, we used two different datasets from different institutes. Institutional review board approval was obtained for this study from both institutions and the need for written informed patient consent was waived. 

\begin{enumerate}
  \item Our source domain contains DWI of 4 b-values; $[b0, b400, b1000, b1600]$ for a total of 533 patients (total of 30616 images) from Sunnybrook Health Sciences Centre, Toronto, Ontario, Canada. The data is split into Training, Validation, and Test sets with a ratio of $3-1-1.5$, respectively. Pixel spacing and the slice thickness of the source domain are respectively 1.46 $\times$ 1.46 millimeters and 3.0 millimeters.  
  \item Our target domain includes DWI containing $[b100]$ of 148 patients (total of 3514 images) from University Health Network, Toronto, Ontario, Canada. Image distortion is a common problem for DWI, for example in the presence of metal artifacts (hip prothesis) or rectal gas, limiting its use in clinical practice, sometimes even making a diagnosis impossible. This problem applies both to radiologists visually assessing a prostate mp-MRI and a CAD tool. Our trained radiologists who performed the segmentations also assessed all imaging studies for sufficient quality. If the quality of any sequence was considered insufficient for clinical assessment, the cases were excluded.
  \par Through our training, we explore different subset sizes of our target domain and the effects of the size of dataset used to fine-tune the transfer learning architecture on the accuracy of prostate WG and TZ segmentation evaluated on a randomly fixed subset of 33 patients (total of 746 images) from the target domain (out of the 148). The pixel spacing and slice thickness of the source domain are respectively 1.66 $\times$ 1.66 millimeters and 3.0 millimeters. Segmentations were performed by a fellowship-trained radiologist with 6 years of experience and a Radiology Resident with 4 years of experience reading prostate mpMRI.
\end{enumerate}

\subsection{Architecture and Training}

\begin{figure*}
\centering
\includegraphics[scale=0.4]{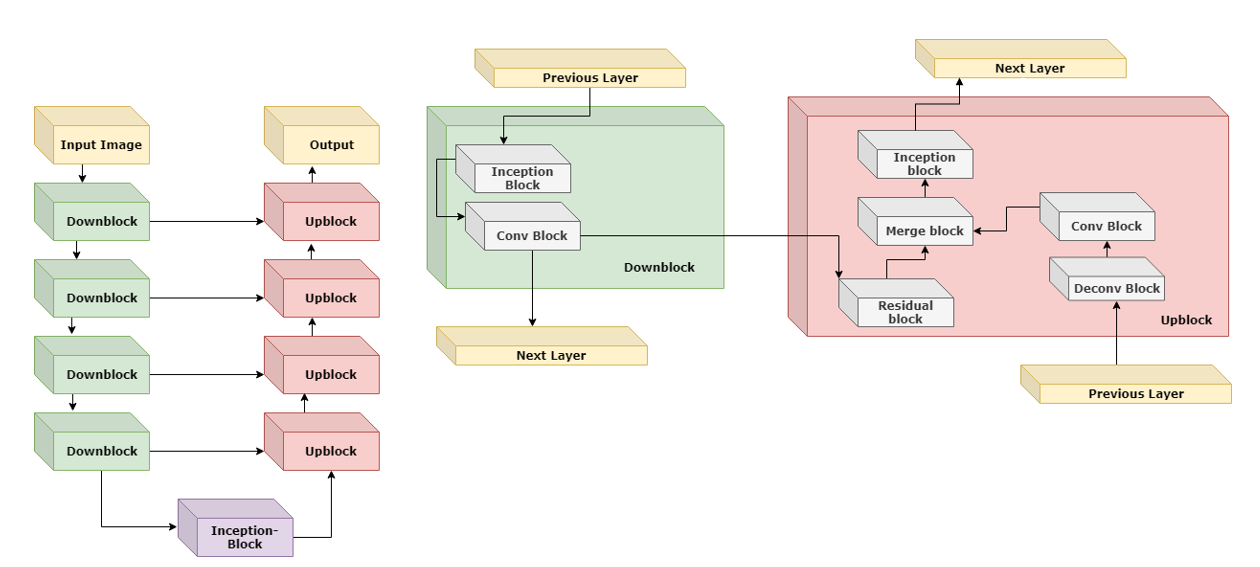}
\caption{Modified U-net Architecture}
\label{fig2}
\smallskip
Left: The modified U-net architecture, which is comprised of Down and Up blocks. Right: Expanded Down and Up blocks.
\end{figure*}

We follow the network architecture proposed by Clark et. al~\cite{1}. Fig~\ref{fig2} shows the proposed architecture where regular convolution blocks in U-net are embedded with inception and residual blocks. Inception blocks apply four convolution and pooling operations in parallel and then concatenate the features at the end of the inception block. Rather than trying different convolution with multiple filter sizes and choosing the best fitting size for the task at hand, inception blocks operate with multiple kernel sizes and combine the results. This has shown to increase segmentation and classification accuracy~\cite{1, 7}. Residual connections were added to the connections between up and down blocks. This allows training deeper networks by minimizing loss of information after each convolution layer, ultimately resulting in reduced areas of false positive \cite{1}. Adam optimizer with the same set of initial learning rates were used for training all the instances of U-net for 25 epochs, using an Nvidia GeForce GTX 1080 Ti 11GB GPU. 

\par In training our base model using the source dataset, we combined the 4 available b-value DW images to increase the number of training images rather than individually training the U-net on each b-value separately and picking the best result. Along with histogram equalization, data augmentation was done using Keras built-in augmentation that performs random horizontal and vertical flips and axis rotations.

\par In convolutional neural networks, shallow layers / features of the network are known to contain more generic features (e.g. edges and size of the object) but deeper layers of the network become progressively more specific to the details of the classes contained in the trained dataset. For the U-net architecture shown in Fig~\ref{fig2}, we expect shallow layers (Down-blocks) to be responsible for learning lower level features which are shared by both source and target datasets, while the deeper layers (Up-blocks) are responsible for more high level features that might differ between the two datasets.

\par After performing histogram equalization on the target domain images, we experimented with different combinations of which layers to fine-tune. Within the radiology community, there is higher agreement on Prostate WG  segmentation compared to that of TZ due to the general shape and vague boundaries of TZ. As a result, for the task of prostate WG segmentation, our best performing model was achieved by fine-tuning only the Up-blocks while keeping the Down-block learned features the same as our source domain learned model. As TZ contouring has more variance between different radiologists compared to WG, our best model for TZ segmentation was achieved by fine-tuning Up-blocks while also fine-tuning the first half of the Down-block layers. We use both conventional and our proposed modified DSC loss function in order to compare the results between and within models (pretrained and transfer learning models). Target domain dataset used for fine-tuning was gradually increased with 10\% increments, starting from 8 patients (194 images) to 115 patients (2,768 images), in order to explore minimum dataset size required to attain acceptable results. The final test dataset from the target domain was kept fixed at 33 patients (746 images) and separate from the fine-tuning datasets.

An overview of the models used for training and comparison purposes are listed below.
    \begin{enumerate}
   \item Training our U-net based architecture from scratch on the target data with conventional DSC loss function (Equation~\ref{eq:2}) ignoring the source domain dataset. 
  \item Transfer Learning by using conventional DSC loss function where we use the features learned from the source domain and use the target domain images to fine-tune and test our model.
   \item Transfer Learning using our proposed modified DSC loss (Equation~\ref{eq:4}). 
   
\end{enumerate}
\subsection{Modified Dice Score Measure}
DSC is the well-accepted measure of segmentation accuracy for medical images in well-known competitions such as PROMISE12~\cite{promise12}. DSC has also been widely used as a loss function for training deep CNNs for the task of medical image segmentation including prostate MRI  ~\cite{1, 103} . DSC measures the overlap between the predicted mask $P$ and ground truth $G$.
\begin{equation}
\label{eq:1}
\textnormal{DSC} = 2 * \dfrac {|P \cap G|+ \epsilon} {|P| + |G|+ \epsilon} 
\end{equation}
\begin{equation}
\label{eq:2}
  \textnormal{Dice-Loss}=
    -1 *  \textnormal{DSC}
\end{equation}
where $\epsilon$ is a small number to avoid division by zero.

By using the DSC loss function defined above, the calculated DSC for cases that do not contain the prostate will be 1. In other words, the reward of predicting a perfect segmentation for prostate containing images and not returning any segmentation prediction for images that do not contain the prostate is the same.

We propose a change to calculating segmentation accuracy by making the following modification to the DSC based loss function where we explore $X$ values from 0.0 to 1.0 with 0.1 increments. 
\begin{equation}
\label{eq:3}
  \textnormal{Modified-DSC}=\begin{cases}
    X & \text{if $G = P = 0$}\\
    2 * \dfrac {|P \cap G|} {|P| + |G|}, & \text{otherwise}.
  \end{cases}
\end{equation}

\begin{equation}
\label{eq:4}
  \textnormal{Modified Dice-Loss}=
    -1 *  \textnormal{Modified-DSC}
\end{equation}
With this modification, we anticipate lower X values to increase segmentation accuracy of the model where training data size is smaller, and bigger X values to perform better in bigger datasets. The reason lies behind the nature of the prostate data where usually $\sim\%50$ of images at patient level are non-prostate images (as it was the case in our dataset) hence by adjusting the reward system in smaller dataset sizes and penalizing correct performance of the model on such images when no segmentation is produced, we shift the focus of the model to the images containing prostate. As training data size increases, the effect of the modified DSC may become negligible since the network will have seen enough cases to reward/penalize correctly.

\subsection{Post-processing}
\begin{figure}
    \centering
    \subfloat[Pre-closing operation]{{\includegraphics[scale=0.67]{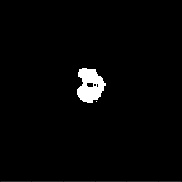} }}%
    \qquad
    \subfloat[Post closing operation]{{\includegraphics[scale=0.67]{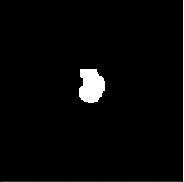} }}%
    \qquad
    \subfloat[Pre-opening operation]{{\includegraphics[scale=0.67]{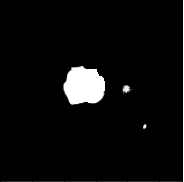} }}%
    \qquad
    \subfloat[Post opening operation]{{\includegraphics[scale=0.5]{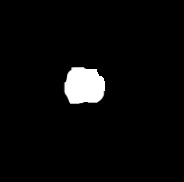} }}%
    \caption{Postprocessing segmentation results}%
    \label{fig3}%
\end{figure}

Two morphological transformations~\cite{Gonzalez2007}, \textit{Opening} and \textit{Closing}, were used in order to improve segmentation accuracy and reduce noise. We know prostate zones are continuous in volume. Closing operation which is a dilation followed by an erosion, fills out holes in the predicted masks by our algorithm. The opening operator which is an erosion followed by dilation, is useful in removing noise outside of our predicted masks. The application of these transformations is shown in Fig~\ref{fig3}. Although DSC improvement with these transformations was minimal ($1\%$), for applications of automated segmentation such as prostate volume calculation, it makes a great difference to have accurate enclosed continuous masks.

The segmentation of prostate WG at the prostate base and apex is challenging for both radiologists and CAD tools. As a result, to mitigate the number of false positive cases in our segmentation task, we dismiss predicted masks below a threshold of 120 pixels which translates to $90\%$ of the mean number of pixels for prostate base and apex (prostate end-points with smallest size within prostate) among all patients. We apply the same methodology to TZ segmentation and filter results at 65 pixels, 90\% of the mean number of pixels for prostate TZ base and apex among all patients. This showed an $1\%$ improvement Dice score in our final result, without sacrificing specificity, sensitivity or precision.

\section{Results}
In the following, the results of the proposed segmentation model is presented for source domain and target domain datasets for both prostate WG and TZ.

\subsection{Source Domain Segmentation Results for Prostate Whole Gland and Transitional zone }
The result of the modified U-net architecture, trained and validated on 448 patients and tested on 85 patients (6,688 images), all from the source domain dataset, can be seen in Table~\ref{table1}. We used our proposed modified DSC loss in training and we report the conventional DSC on the test set images containing prostate (images without prostate are ignored). We calculated specificity, sensitivity, and precision for detection of all patient images, including and excluding the prostate.
\begin{table}
\centering
\caption{Segmentation result for source domain dataset}
 \begin{tabular}{||c c c c c||} 
 \hline
 Zone & DSC & Specificity & Sensitivity & Precision \\ [0.5ex] 
 \hline\hline
 WG & $0.89 \pm 0.01$ & $0.97$ & $0.96$ & $0.98$ \\ 
 \hline
 TZ & $0.86 \pm 0.02$ & $0.97$ & $0.96$ & $0.98$ \\ 
 \hline
 \end{tabular}
\medskip
\label{table1}
\end{table}

\subsection{Target Domain Segmentation}
From the target domain, we randomly picked 33 patients (20\% of total number of images) as our test set through all iterations of training. For fine-tuning the model, we used different sizes from 8 patients to 115 patients in the target dataset. Unlike training source domain data, in training (fine-tuning) the target domain images, we did not use data augmentation to keep the training fast and explore minimal data size requirement for transfer learning to achieve optimal segmentation accuracy.

\par For each set target size, we trained our modified U-net in 2 different ways; 1) Training the U-net from scratch with exploring DSC loss modifications, using target domain training data only 2) Transfer Learning by using modifications to the DSC loss, pretrained on source domain and fine-tuned on target domain. We also used the performance of our trained network on source domain and applied, without any training (fine-tuning), to the target domain as a baseline. Source domain's performance on its test set acts as an optimal performing result for comparison purposes.

\subsubsection{Prostate WG}
Table~\ref{table2} shows detailed average DSCs over training with different training data size in the target domain. ``Transfer Learning" and "Train Target from Scratch" are the results of our training with and without a transfer learning approach after exploring different $X$ values from Equation~\ref{eq:3}. We achieved higher accuracy with $X = 0$ for transfer learning on all dataset sizes and for training target from scratch only for dataset sizes of 8 and 30 patients. For the rest of the experiments with data sizes of 42 patients and higher (42-115), the conventional loss where $X = 1.0$ results in a higher accuracy. Hence, the reported DSC for both transfer learning and training from scratch uses our modified DSC loss, except for training from scratch on data sizes of 42-115 patients. Transfer learning DSC results with our proposed DSC loss performed, on average, 1\% better compared to DSC achieved with conventional loss. Our proposed loss performed much better in dataset sizes of 8 and 30 when the model was trained from scratch. We achieved A DSC of 0.58 (confidence interval CI = [56, 59]) compared to 0.44 (standard deviation \(\sigma = 0.02\), CI = [0.42,0.45]),  without overlap in confidence interval between the two, when training on 8 patients and 0.65 (CI =[63, 66])  compared to 0.57 (\(\sigma = 0.01\), CI = [56, 57]), without overlap in confidence interval between the two, on 30 patients. ``No Training on Target" uses the model trained only using the source data and applies it to the target domain test cases.
\begin{table}
\centering
\caption{DSC results for segmentation of prostate WG in the target domain}
 \begin{tabular}{||c c c c||}
 \hline
 Target Size & Transfer & Train Target & No Training\\ [0.5ex]
             & Learning & from Scratch & on Target\\ [0.5ex] 
 \hline\hline
 8  &  $0.79\pm 0.01$ & $0.58\pm 0.02$  & 0.64\\ 
 \hline
 30   & $0.8\pm 0.01$ & $0.65\pm 0.02$ & 0.64 \\ 
  \hline
 42   & $0.81\pm 0.01$ & $0.68\pm 0.02$ & 0.64\\  
  \hline
  70   & $0.82\pm 0.01$ & $0.73\pm 0.01$ & 0.64\\  
  \hline 
  85   & $0.83\pm 0.01$ & $0.76\pm 0.01$ & 0.64\\  
  \hline
  106   & $0.84\pm 0.01$ & $0.77\pm 0.01$ & 0.64\\  
  \hline
  115   & $0.85\pm 0.01$ & $0.79\pm 0.01$ & 0.64\\  
  \hline
\end{tabular}
\smallskip
\label{table2}
\end{table}

\begin{figure}
\centering
\includegraphics[scale=0.55]{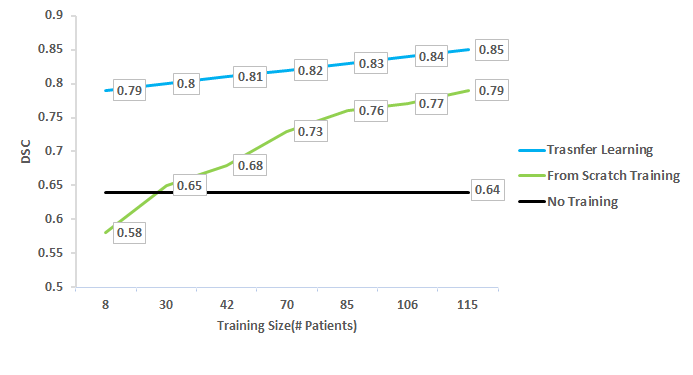}
\caption{DSC trend on segmentation of the target domain prostate WG}
\medskip
\label{fig4}
\end{figure}

Fig~\ref{fig4} shows the improvement of average DSCs for WG on our 33 test patients, trained with and without transfer learning and modified DSC loss. The \textbf{blue} line shows the result of transfer learning with modified DSC where $X = 0$ in Equation~\ref{eq:3} and our model is pretrained using the source data and fine-tuned using the target data with different sizes (8-115). We used our modified loss on dataset sizes of 8 and 30 for training the model from scratch (\textbf{green} line). For n = 42-115, we used the conventional loss. the \textbf{black} line shows the result of applying the pretrained model to the 33 target test patients, without any training.

\par
\begin{figure}
    \centering
    \subfloat[8 Patients]{{\includegraphics[scale=0.45]{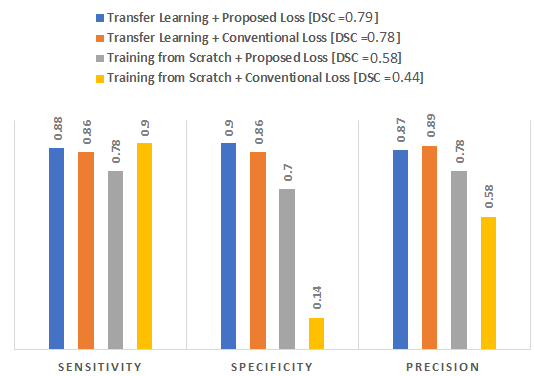} }}%
    \qquad
    \subfloat[70 Patients]{{\includegraphics[scale=0.45]{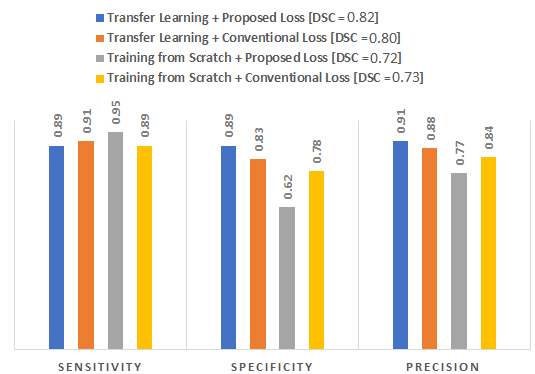} }}%
     \qquad
    \subfloat[115 Patients]{{\includegraphics[scale=0.45]{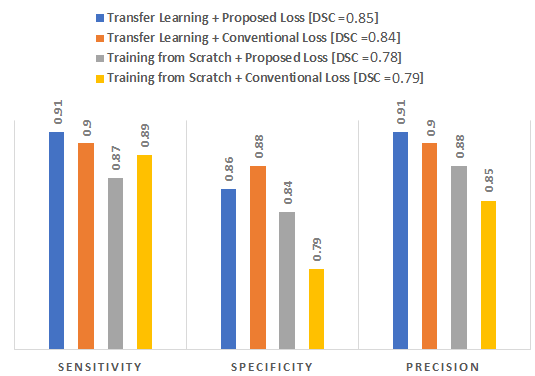} }}%
    \caption{The overall sensitivity, specificity and precision for WG detection using modified DSC loss, both with a transfer learning and from scratch training, trained with 8, 70 and 115 patients and tested on 33 patients.}%
\label{fig5}
\end{figure}

 Transfer learning is able to extend our WG segmenting model, trained on source data to target data with a DSC of $\sim{0.79}$ with only 8 cases from the target dataset that are contoured by the radiologist. Training the U-net from scratch will result in a DSC of $\sim{0.58}$ with a longer training time compared to the transfer learning approach. Even though our modified DSC loss achieves a higher overall DSC compared to the conventional DSC counterpart in the smaller dataset sizes, consideration of measures such as specificity, sensitivity and precision on image level show that DSC alone is not a good measure of accuracy for segmentation tasks. Fig~\ref{fig5} compares sensitivity, specificity and precision of testing our model, trained with 8, 70 and 115 patients using the conventional DSC loss measured against our proposed modified Dice loss.

The proposed DSC loss, where $X = 0$ in Equation~\ref{eq:3}, and its counterpart conventional loss both performed well using a transfer learning approach, with DSC, sensitivity, specificity and precision not being persistently better in one or the other. The proposed loss however performs much better when training a model from scratch, using a limited training dataset as can be seen in Fig~\ref{fig5} (a).

\begin{figure}
    \begin{minipage}[t]{.22\textwidth}
        \centering
        \includegraphics[width=\textwidth]{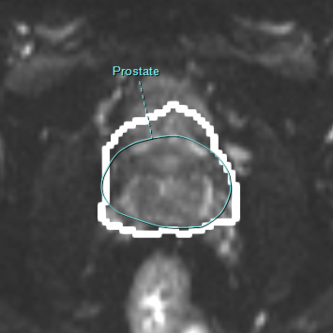}
        \caption{(a) From Scratch }\label{fig:1}
    \end{minipage}
    \hfill
    \begin{minipage}[t]{.22\textwidth}
        \centering
        \includegraphics[width=\textwidth]{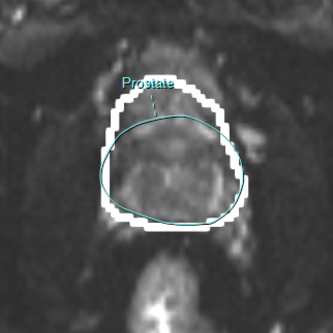}
        \caption{(b) From Scratch + Modified DSC loss}\label{fig:2}
    \end{minipage}
    \hfill
    \begin{minipage}[t]{.22\textwidth}
        \centering
        \includegraphics[width=\textwidth]{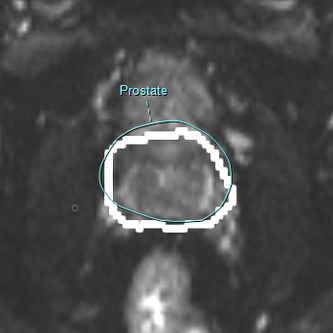}
        \caption{(c) Transfer Learning}\label{fig:3}
    \end{minipage}  
    \hfill
    \begin{minipage}[t]{.22\textwidth}
        \centering
        \includegraphics[width=\textwidth]{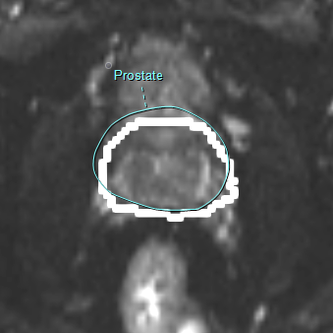}
        \caption{(d) Transfer Learning + Modified DSC loss }\label{fig:2}
    \end{minipage}
   
    \caption{Predicted prostate WG mask (white) vs. ground truth (blue) for sample cases. From left to write, \textbf{a)} Training from scratch \textbf{b)} Training from Scratch with modified DSC \textbf{c)} Transfer Learning \textbf{d)} Transfer Learning with modified DSC.}
    \label{fig6}
\end{figure}

Fig~\ref{fig6} shows the prediction of prostate WG compared to the ground truth contoured by radiologists, using our loss function against its counterpart. As expected, when training from scratch, the conventional loss performs better than our proposed loss. The subtle improvement of predictions as we move from training from scratch and conventional DSC to our transfer learning approach with modified DSC may not be visible in the figure.

\subsubsection{Prostate TZ}
For prostate TZ segmentation, we followed the same training scheme as for WG, keeping our test patients unchanged. Table~\ref{table3} shows detailed average DSCs over training with different sizes of target domain data.

\begin{table}
\centering
\caption{DSC results for segmentation of prostate TZ in the target domain}
 \begin{tabular}{||c c c c ||} 
 \hline
Target Size & Transfer & Train Target & No Training\\ [0.5ex]
             & Learning & from Scratch & on Target\\ [0.5ex] 
 
 \hline\hline
 8   & $0.77\pm 0.01$ & $0.39\pm 0.02$  & 0.71\\ 
 \hline
 30  & $0.80\pm 0.01$ & $0.51\pm 0.01$  &  0.71 \\ 
  \hline
 42   & $0.80 \pm 0.01$ & $0.67\pm 0.01$ &   0.71\\  
  \hline
  70  & $0.81\pm 0.01$ & $0.71\pm 0.01$ &   0.71\\  
  \hline 
  85   & $0.82\pm 0.01$ & $0.74\pm 0.01$ &  0.71\\  
  \hline
  106   & $0.82\pm 0.01$ & $0.77\pm 0.01$ &   0.71\\  
  \hline
  115  & $0.84\pm 0.01$ & $0.79\pm 0.01$ &   0.71\\  
  \hline
\end{tabular}
\smallskip
\label{table3}
\end{table}
\par
\begin{figure}

\centering
\includegraphics[scale=0.55]{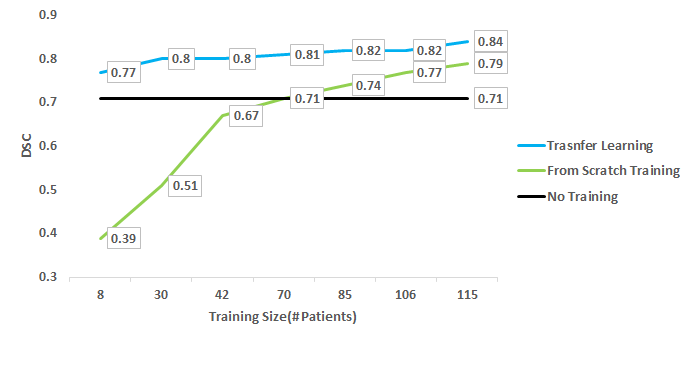}
\caption{DSC trend on segmentation of the target domain prostate TZ}
\label{fig7}
\end{figure}

 Fig~\ref{fig7} shows the improvement of average DSCs for TZ on our 33 test patients, trained with and without transfer learning and modified DSC loss. The \textbf{blue} line which is the result of transfer learning uses our modified DSC loss on all training instances (n = 8-115) while we used the same methodology as prostate WG for training from scratch and used the modified DSC loss on 8 and 30 patient data sizes and used conventional loss for n = 42-115.

\par
\begin{figure}[htb]
    \begin{minipage}[t]{.22\textwidth}
        \centering
        \includegraphics[width=\textwidth]{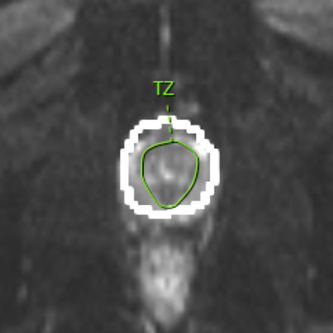}
        \caption{(a) From Scratch }
    \end{minipage}
    \hfill
    \begin{minipage}[t]{.22\textwidth}
        \centering
        \includegraphics[width=\textwidth]{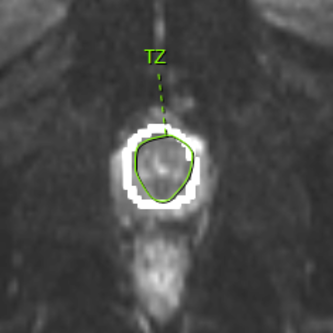}
        \caption{(b) From Scratch + Modified DSC loss}
    \end{minipage}
    \hfill
    \begin{minipage}[t]{.22\textwidth}
        \centering
        \includegraphics[width=\textwidth]{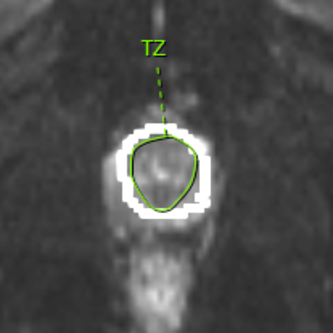}
        \caption{(c) Transfer Learning}
    \end{minipage}  
    \hfill
    \begin{minipage}[t]{.22\textwidth}
        \centering
        \includegraphics[width=\textwidth]{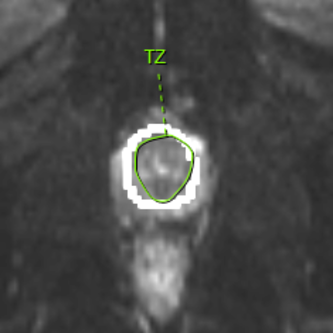}
        \caption{(d) Transfer Learning + Modified DSC loss }
    \end{minipage}
    
    \caption{Predicted prostate TZ (white) vs. ground truth (blue) for sample cases. From left to write, \textbf{a)} Training from scratch \textbf{b)} Training from Scratch with modified DSC \textbf{c)} Transfer Learning \textbf{d)} Transfer Learning with modified DSC}
    \label{fig8}
\end{figure}

 Fig~\ref{fig8} shows the prediction of prostate TZ compared to the ground truth of the zone contoured by radiologists, using our loss function against its counterpart. 

\section{Discussion}
With the increase in using DWI for the purpose of prostate cancer diagnosis and prognosis, automatic accurate segmentation of DWI into prostate WG and TZ can assist radiologists in diagnosis and prognosis of prostate cancer. In this work, we have proposed a transfer learning architecture to extend a trained CNN model on a large MRI dataset, to target domain MRI images with different acquisition parameters. With the limitations on available data size in the field of medical imaging, transfer learning allows the use of all available data from different cohorts for a task at hand. As shown in our results, training on the source domain and crudely applying the trained model to target domain results in poor performance (DSCs of 0.64 and 0.71 for WG and TZ, respectively). Moreover, training a model from scratch never reaches the accuracy of the proposed transfer learning method regardless of the size of the training dataset in the target cohort (Fig~\ref{fig4} and Fig~\ref{fig7}). To find the best combination of layers to fine-tune for the task of segmentation based on our data, we experimented with 10$\%$ (in terms of number of layers) increases, in direction of both ends of the network.For the task of prostate WG segmentation, our best performing model was achieved by fine-tuning only the Up-blocks while keeping the Down-block learned features the same as our source domain learned model. Our best model for TZ segmentation was achieved by fine-tuning Up-blocks while also fine-tuning the first half of the Down-block layers.

The goal of this work was also to identify minimum required labeled dataset size, in order to achieve acceptable segmentation results. To justify the need for a transfer learning approach for the task of segmentation when multiple cohorts are present, we had a training instance with images from both source and target domain combined for training and we kept the 30 test patients the same. The average DSC of 5 such training instances was 0.81 which under-performed when training via transfer learning. We have shown that with datasets as small as 30 patients, we can extend our model, trained on the source domain containing 533 patients, to a target domain with DSC of $\sim{0.80}$ on the test set for segmentation of both prostate WG and  TZ segmentation. As the number of training cases in the target domain increases (115), the DSC of prostate WG and TZ segmentation reaches to 0.85 and 0.84, respectively. The improvements are statistically significant when using 115 patients compared to 30 patients, with no overlap in confidence intervals (for prostate WG, with 30 patients the CI is [0.78, 0.79] and for 115 patients, the CI is [0.84, 0.85]). This is a reasonable DSC, which can be accepted for different applications such as prostate volume calculation. Given the small training dataset required on the target domain, the proposed approach can have significant practical impact.

\par We also proposed a loss function based on a modified DSC measure. Although the proposed loss function does not show statistically significant improvement in overall DSC on the test images while using the transfer learning approach, it shows great improvement on DSC, specificity, and precision of detecting images with prostate when training the network from scratch using with small training datasets. This was achieved by making the network focus only on images containing prostate. As the dataset size grows, training from scratch with this method leads to lower specificity with the increase of false positives. Another implication of our modified DSC loss function is showing that DSC, on its own, as currently used by researchers, may not be the best measure for reporting accuracy. As we showed in Fig~\ref{fig5}, DSC becomes more meaningful when combined with sensitivity, specificity and precision measures to give a bigger picture of accuracy of different models (e.g. Fig~\ref{fig5} (c)).

Some of the challenges and limitations of prostate segmentation is the ambiguity and difficulty in segmentation of base and apex regions by radiologists. Our best performing model for prostate WG segmentation results in 50 false negative and positive predictions (0.06\% of total images in test set). Out of these mis-predictions, we estimated what portion belong to apex/base and what portion belong to mid-gland. We used pixel thresholding based on min / max of appearing images (base/apex masks are much smaller than mid-gland on average) and positions of images (if there are predictions before and after a slice, it is most-likely mid-gland and base/apex otherwise). Only around 14 images (28\%) of the mis-predictions were not base/apex while the rest belong to either base or apex of prostate. Out of those 36 images, different radiologists may or may not include those images as a part of the prostate since the zones get small and difficult to detect. A future work could make use of robust registration / detection methods at the base and apex to improve the segmentation of these slices.

This work has limitations. We did not take inter-observer variability in prostate segmentation into account. This can be a future work to measure the effect of such variability in developing training data and on the robustness of the segmentation results in the test set. In addition, we did not calculate prostate volume based on segmentation results. Prostate volume may be a prognostic factor for prostate cancer and in future work, the performance of the proposed segmentation algorithm for calculating prostate volume can be evaluated.

\section{Conclusion}
\par This work explores different transfer learning approaches for prostate WG and TZ segmentation by extending one cohort to another with a small amount of labeled data. This will allow clinicians and radiologists to have a guideline for the optimal number of contoured images they need in order to successfully achieve segmentation results based on their accuracy needs. The modification of Dice coefficient loss eliminated the need for training a separate network for detecting the prostate before performing segmentation, while outperforming the conventional Dice loss. This model can be used to calculate prostate WG and TZ volumes and to preprocess MR images in order to perform anomaly detection computer aided models that require the segmentation of prostate WG and TZ.

\vspace{2ex}\noindent\textbf{Biographies} 

\vspace{2ex}\noindent\textbf{Saman Motamed} received his HBSc degree in Computer Science from University of Toronto, Toronto, Canada in 2019. He is currently pursuing a MSc in Medical Science, at the University of Toronto, Toronto, Canada.
His research interest include development of Machine Learning, and Computer Vision architectures with focus on Medical Imaging.
\vspace{2ex}\noindent\textbf{Dr. Isha Gujrathi}  is a qualified radiologist trained in India and is presently working as a Postdoc Research Fellow at the Brigham and Women's Hospital in Boston, Masahcussetts, USA. Her research focus is biomedical and imaging informatics and she is actively involved in data preparation and analysis. Her interests pertain to use of evidence-based medicine, radiomics and informatics to make imaging more efficient and improve the quality of radiology reporting.

\vspace{2ex}\noindent\textbf{Dr. Dominik Deniffel} has completed 4 years of radiology residency training, with his most recent appointment at the Technical University of Munich, Germany. He has held research fellow positions at the University of California San Diego, USA, the University of Strasbourg, France and the Technical University of Munich, Germany. He is currently working at the Lunenfeld-Tanenbaum Research Institute in Toronto, Canada.\\ His main research interests are quantitative cancer imaging, with a focus on genitourinary cancers, and implementation of artificial intelligence applications into clinical practice.

\vspace{2ex}\noindent\textbf{Dr. Anton Oentoro}
Bachelor (Hons.) of Computing Science
in Biomedical Computing in 2007, as well
as a Master’s Degree of Science
at Queen’s University in 2009 in Kingston, Canada. He went on to complete his
Medical Degree at the University of
Toronto (2013) and completed his
residency training in Diagnostic
Radiology from the same institution
(2018). Subsequently, he
completed additional fellowship training
in Abdominal Radiology at the University
of Toronto (2019).

\vspace{2ex}\noindent\textbf{Dr. Masoom A. Haider} completed his medical training at the University of Ottawa, Ottawa, Canada and University of Toronto, Toronto, Canada and the Cleveland Clinic in Abdominal Imaging and Body MRI. As the Head of Abdominal MRI until 2013 he built the largest Abdominal MRI program in Canada and pioneered the use of multiparametric MRI in Canada to detect prostate cancer, a technique now used around the world. He has over 180 publications including multiple publications in the fields of radiomics and machine learning and has held multiple peer reviewed grants.

\vspace{2ex}\noindent\textbf{Dr. Farzad Khalvati} received his Ph.D. in Electrical and Computer Engineering at the University of Waterloo, Waterloo, Ontario, Canada, in 2009. He is currently an Assistant Professor in the Department of Medical Imaging with cross-appointment to the Department of Mechanical and Industrial Engineering at the University of Toronto. His research expertise and interests are multidisciplinary and include artificial intelligence and computer vision with focus on medical imaging, translational medicine, and clinical decision support systems.

\vspace{2ex}\noindent\textbf{Authors' contributions}
SM, MAH, and FK contributed to the design of the concept. IG, DD, AO, and MAH contributed in collecting and reviewing the data. SM and FK contributed to the design and implementation of machine learning modules. All authors contributed to the writing and reviewing of the paper. FK and MAH are co-senior authors. All authors read and approved the final manuscript. 

\vspace{2ex}\noindent\textbf{Disclosures}
The authors have no competing interests to declare.

\vspace{2ex}\noindent\textbf{Funding Acknowledgment}
This research was conducted with the support of the Ontario Institute for Cancer Research (OICR) through funding provided by the Government of Ontario.

\bibliography{references}

\begin{thebibliography}{10}

\bibitem{39}
Freddie Bray, Jacques Ferlay, Isabelle Soerjomataram, Rebecca~L Siegel,
  Lindsey~A Torre, and Ahmedin Jemal.
\newblock Global cancer statistics 2018: Globocan estimates of incidence and
  mortality worldwide for 36 cancers in 185 countries.
\newblock {\em CA: a cancer journal for clinicians}, 68(6):394--424, 2018.

\bibitem{41}
Alessandro Sciarra, Jelle Barentsz, Anders Bjartell, James Eastham, Hedvig
  Hricak, Valeria Panebianco, and J~Alfred Witjes.
\newblock Advances in magnetic resonance imaging: how they are changing the
  management of prostate cancer.
\newblock {\em European urology}, 59(6):962--977, 2011.

\bibitem{Roobol2012}
Monique~J. Roobol, Heidi~A. {Van Vugt}, Stacy Loeb, Xiaoye Zhu, Meelan Bul,
  Chris~H. Bangma, Arno~G.L.J.H. {Van Leenders}, Ewout~W. Steyerberg, and
  Fritz~H. Schr{\"{o}}der.
\newblock {Prediction of prostate cancer risk: The role of prostate volume and
  digital rectal examination in the ERSPC risk calculators}.
\newblock {\em European Urology}, 2012.

\bibitem{Cameron2016}
Andrew Cameron, Farzad Khalvati, Masoom Haider, and Alexander Wong.
\newblock {MAPS: A Quantitative Radiomics Approach for Prostate Cancer
  Detection.}
\newblock {\em IEEE transactions on bio-medical engineering}, 63(6):1145--1156,
  2016.

\bibitem{Chung2015}
A.G. Chung, F.~Khalvati, M.J. Shafiee, M.A. Haider, and A.~Wong.
\newblock {Prostate cancer detection via a quantitative radiomics-driven
  conditional random field framework}.
\newblock {\em IEEE Access}, 3, 2015.

\bibitem{30}
Farzad Khalvati, Alexander Wong, and Masoom~A Haider.
\newblock Automated prostate cancer detection via comprehensive
  multi-parametric magnetic resonance imaging texture feature models.
\newblock {\em BMC medical imaging}, 15(1):27, 2015.

\bibitem{Khalvati2018}
F.~Khalvati, J.~Zhang, A.G. Chung, M.J. Shafiee, A.~Wong, and M.A. Haider.
\newblock {MPCaD: A multi-scale radiomics-driven framework for automated
  prostate cancer localization and detection}.
\newblock {\em BMC Medical Imaging}, 18(1), 2018.

\bibitem{29}
Geert Litjens, Robert Toth, Wendy van~de Ven, Caroline Hoeks, Sjoerd Kerkstra,
  Bram van Ginneken, Graham Vincent, Gwenael Guillard, Neil Birbeck, Jindang
  Zhang, et~al.
\newblock Evaluation of prostate segmentation algorithms for mri: the promise12
  challenge.
\newblock {\em Medical image analysis}, 18(2):359--373, 2014.

\bibitem{28}
Shijun Wang, Karen Burtt, Baris Turkbey, Peter Choyke, and Ronald~M Summers.
\newblock Computer aided-diagnosis of prostate cancer on multiparametric mri: a
  technical review of current research.
\newblock {\em BioMed research international}, 2014, 2014.

\bibitem{DBLP:journals/corr/abs-1905-13145}
Sunghwan Yoo, Isha Gujrathi, Masoom~A. Haider, and Farzad Khalvati.
\newblock Prostate cancer detection using deep convolutional neural networks.
\newblock {\em arXiv}, abs/1905.13145, 2019.

\bibitem{meyer2019towards}
Anneke Meyer, Marko Rakr, Daniel Schindele, Simon Blaschke, Martin Schostak,
  Andriy Fedorov, and Christian Hansen.
\newblock Towards patient-individual pi-rads v2 sector map: Cnn for automatic
  segmentation of prostatic zones from t2-weighted mri.
\newblock In {\em 2019 IEEE 16th International Symposium on Biomedical Imaging
  (ISBI 2019)}, pages 696--700. IEEE, 2019.

\bibitem{jensen2019prostate}
Carina Jensen, Kristine~Storm S{\o}rensen, Cecilia~Klitgaard J{\o}rgensen,
  Camilla~Winther Nielsen, Pia~Christine H{\o}y, Niels~Christian Langkilde, and
  Lasse~Riis {\O}stergaard.
\newblock Prostate zonal segmentation in 1.5 t and 3t t2w mri using a
  convolutional neural network.
\newblock {\em Journal of Medical Imaging}, 6(1):014501, 2019.

\bibitem{3}
Olaf Ronneberger, Philipp Fischer, and Thomas Brox.
\newblock U-net: Convolutional networks for biomedical image segmentation.
\newblock pages 234--241, 2015.

\bibitem{2}
Md~Zahangir Alom, Mahmudul Hasan, Chris Yakopcic, Tarek~M Taha, and Vijayan~K
  Asari.
\newblock Recurrent residual convolutional neural network based on u-net
  (r2u-net) for medical image segmentation.
\newblock {\em arXiv preprint arXiv:1802.06955}, 2018.

\bibitem{102}
Bashar Zelhof, Martin Pickles, Gary Liney, Peter Gibbs, Greta Rodrigues, Sigurd
  Kraus, and Lindsay Turnbull.
\newblock Correlation of diffusion-weighted magnetic resonance data with
  cellularity in prostate cancer.
\newblock {\em BJU international}, 103(7):883--888, 2009.

\bibitem{100}
Masoom~A Haider, Theodorus~H Van Der~Kwast, Jeff Tanguay, Andrew~J Evans,
  Ali-Tahir Hashmi, Gina Lockwood, and John Trachtenberg.
\newblock Combined t2-weighted and diffusion-weighted mri for localization of
  prostate cancer.
\newblock {\em American journal of roentgenology}, 189(2):323--328, 2007.

\bibitem{101}
Jeffrey~C Weinreb, Jelle~O Barentsz, Peter~L Choyke, Francois Cornud, Masoom~A
  Haider, Katarzyna~J Macura, Daniel Margolis, Mitchell~D Schnall, Faina
  Shtern, Clare~M Tempany, et~al.
\newblock Pi-rads prostate imaging--reporting and data system: 2015, version 2.
\newblock {\em European urology}, 69(1):16--40, 2016.

\bibitem{38}
Roland Bammer.
\newblock Basic principles of diffusion-weighted imaging.
\newblock {\em European journal of radiology}, 45(3):169--184, 2003.

\bibitem{42}
Sinno~Jialin Pan and Qiang Yang.
\newblock A survey on transfer learning.
\newblock {\em IEEE Transactions on knowledge and data engineering},
  22(10):1345--1359, 2009.

\bibitem{4}
Mohsen Ghafoorian, Alireza Mehrtash, Tina Kapur, Nico Karssemeijer, Elena
  Marchiori, Mehran Pesteie, Charles~RG Guttmann, Frank-Erik de~Leeuw, Clare~M
  Tempany, Bram van Ginneken, et~al.
\newblock Transfer learning for domain adaptation in mri: Application in brain
  lesion segmentation.
\newblock pages 516--524, 2017.

\bibitem{20}
Chuen-Kai Shie, Chung-Hisang Chuang, Chun-Nan Chou, Meng-Hsi Wu, and Edward~Y
  Chang.
\newblock Transfer representation learning for medical image analysis.
\newblock pages 711--714, 2015.

\bibitem{21}
Maithra Raghu, Chiyuan Zhang, Jon~M. Kleinberg, and Samy Bengio.
\newblock Transfusion: Understanding transfer learning with applications to
  medical imaging.
\newblock {\em CoRR}, abs/1902.07208, 2019.

\bibitem{105}
Lei Zhang, Aly~A Mohamed, Ruimei Chai, Yuan Guo, Bingjie Zheng, and Shandong
  Wu.
\newblock Automated deep learning method for whole-breast segmentation in
  diffusion-weighted breast mri.
\newblock {\em Journal of Magnetic Resonance Imaging}, 51(2):635--643, 2020.

\bibitem{104}
Konstantinos Kamnitsas, Christian Baumgartner, Christian Ledig, Virginia
  Newcombe, Joanna Simpson, Andrew Kane, David Menon, Aditya Nori, Antonio
  Criminisi, Daniel Rueckert, et~al.
\newblock Unsupervised domain adaptation in brain lesion segmentation with
  adversarial networks.
\newblock In {\em International conference on information processing in medical
  imaging}, pages 597--609. Springer, 2017.

\bibitem{8}
Chen Shen, Holger~R Roth, Hirohisa Oda, Masahiro Oda, Yuichiro Hayashi,
  Kazunari Misawa, and Kensaku Mori.
\newblock On the influence of dice loss function in multi-class organ
  segmentation of abdominal ct using 3d fully convolutional networks.
\newblock {\em arXiv preprint arXiv:1801.05912}, 2018.

\bibitem{1}
Tyler Clark, Junjie Zhang, Sameer Baig, Alexander Wong, Masoom~A Haider, and
  Farzad Khalvati.
\newblock Fully automated segmentation of prostate whole gland and transition
  zone in diffusion-weighted mri using convolutional neural networks.
\newblock {\em Journal of Medical Imaging}, 4(4):041307, 2017.

\bibitem{vgg}
Karen Simonyan and Andrew Zisserman.
\newblock Very deep convolutional networks for large-scale image recognition.
\newblock {\em arXiv preprint arXiv:1409.1556}, 2014.

\bibitem{Khalvati2015c}
Farzad Khalvati, Aryan Salmanpour, Shahryar Rahnamayan, Masoom~A. Haider, and
  H.~R. Tizhoosh.
\newblock {Sequential Registration-Based Segmentation of the Prostate Gland in
  MR Image Volumes}.
\newblock {\em Journal of Digital Imaging}, 29:254--263, 2016.

\bibitem{Khalvati2013}
Farzad Khalvati, Aryan Salmanpour, Shahryar Rahnamayan, George Rodrigues, and
  Hamid~R Tizhoosh.
\newblock {Inter-slice Bidirectional Registration-based Segmentation of the
  Prostate Gland in MR and CT Image Sequences}.
\newblock {\em Medical physics}, 40(12):123503--1--11, 2013.

\bibitem{10}
Stefan Klein, Uulke~A Van Der~Heide, Irene~M Lips, Marco Van~Vulpen, Marius
  Staring, and Josien~PW Pluim.
\newblock Automatic segmentation of the prostate in 3d mr images by atlas
  matching using localized mutual information.
\newblock {\em Medical physics}, 35(4):1407--1417, 2008.

\bibitem{11}
S{\'e}bastien Martin, Vincent Daanen, and Jocelyne Troccaz.
\newblock Atlas-based prostate segmentation using an hybrid registration.
\newblock {\em International Journal of Computer Assisted Radiology and
  Surgery}, 3(6):485--492, 2008.

\bibitem{12}
Junjie Zhang, Sameer Baig, Alexander Wong, Masoom~A Haider, and Farzad
  Khalvati.
\newblock Segmentation of prostate in diffusion mr images via clustering.
\newblock pages 471--478, 2017.

\bibitem{18}
Ruida Cheng, Holger~R Roth, Nathan~S Lay, Le~Lu, Baris Turkbey, William
  Gandler, Evan~S McCreedy, Thomas~J Pohida, Peter~A Pinto, Peter~L Choyke,
  et~al.
\newblock Automatic magnetic resonance prostate segmentation by deep learning
  with holistically nested networks.
\newblock {\em Journal of Medical Imaging}, 4(4):041302, 2017.

\bibitem{19}
Xin Yi, Ekta Walia, and Paul Babyn.
\newblock Generative adversarial network in medical imaging: A review.
\newblock {\em arXiv preprint arXiv:1809.07294}, 2018.

\bibitem{31}
Yann LeCun, L{\'e}on Bottou, Yoshua Bengio, Patrick Haffner, et~al.
\newblock Gradient-based learning applied to document recognition.
\newblock {\em Proceedings of the IEEE}, 86(11):2278--2324, 1998.

\bibitem{32}
Alex Krizhevsky, Ilya Sutskever, and Geoffrey~E Hinton.
\newblock Imagenet classification with deep convolutional neural networks.
\newblock In {\em Advances in neural information processing systems}, pages
  1097--1105, 2012.

\bibitem{33}
Pierre Sermanet, David Eigen, Xiang Zhang, Micha{\"e}l Mathieu, Rob Fergus, and
  Yann LeCun.
\newblock Overfeat: Integrated recognition, localization and detection using
  convolutional networks.
\newblock {\em arXiv preprint arXiv:1312.6229}, 2013.

\bibitem{34}
Karen Simonyan and Andrew Zisserman.
\newblock Very deep convolutional networks for large-scale image recognition.
\newblock {\em arXiv preprint arXiv:1409.1556}, 2014.

\bibitem{35}
Christian Szegedy, Wei Liu, Yangqing Jia, Pierre Sermanet, Scott Reed, Dragomir
  Anguelov, Dumitru Erhan, Vincent Vanhoucke, and Andrew Rabinovich.
\newblock Going deeper with convolutions.
\newblock In {\em Proceedings of the IEEE conference on computer vision and
  pattern recognition}, pages 1--9, 2015.

\bibitem{36}
Kaiming He, Xiangyu Zhang, Shaoqing Ren, and Jian Sun.
\newblock Deep residual learning for image recognition.
\newblock In {\em Proceedings of the IEEE conference on computer vision and
  pattern recognition}, pages 770--778, 2016.

\bibitem{22}
{\"O}zg{\"u}n {\c{C}}i{\c{c}}ek, Ahmed Abdulkadir, Soeren~S. Lienkamp, Thomas
  Brox, and Olaf Ronneberger.
\newblock 3d u-net: Learning dense volumetric segmentation from sparse
  annotation.
\newblock pages 424--432, 2016.

\bibitem{7}
Christian Szegedy, Sergey Ioffe, Vincent Vanhoucke, and Alexander~A Alemi.
\newblock Inception-v4, inception-resnet and the impact of residual connections
  on learning.
\newblock 2017.

\bibitem{promise12}
Geert Litjens, Robert Toth, Wendy van~de Ven, Caroline Hoeks, Sjoerd Kerkstra,
  Bram van Ginneken, Graham Vincent, Gwenael Guillard, Neil Birbeck, Jindang
  Zhang, et~al.
\newblock Evaluation of prostate segmentation algorithms for mri: the promise12
  challenge.
\newblock {\em Medical image analysis}, 18(2):359--373, 2014.

\bibitem{103}
Michal Drozdzal, Eugene Vorontsov, Gabriel Chartrand, Samuel Kadoury, and Chris
  Pal.
\newblock The importance of skip connections in biomedical image segmentation.
\newblock In {\em Deep Learning and Data Labeling for Medical Applications},
  pages 179--187. Springer, 2016.

\bibitem{Gonzalez2007}
Rafael~C Gonzalez and Richard~E Woods.
\newblock {\em {Digital Image Processing (3rd Edition)}}.
\newblock 2007.

\end{thebibliography}
\end{document}